# Revisiting the scaling of citations for research assessment[1]

*Giovanni Abramo[a,b], Tindaro Cicero[b], Ciriaco Andrea D'Angelo[b]*

[a] Institute for System Analysis and Computer Science (IASI-CNR)
National Research Council of Italy

[b] Laboratory for Studies of Research and Technology Transfer
School of Engineering, Department of Management
University of Rome "Tor Vergata"

**Abstract**

Over the past decade, national research evaluation exercises, traditionally conducted using the peer review method, have begun opening to bibliometric indicators. The citations received by a publication are assumed as proxy for its quality, but they require standardization prior to use in comparative evaluation of organizations or individual scientists: the citation data must be standardized, due to the varying citation behavior across research fields. The objective of this paper is to compare the effectiveness of the different methods of normalizing citations, in order to provide useful indications to research assessment practitioners. Simulating a typical national research assessment exercise, he analysis is conducted for all subject categories in the hard sciences and is based on the Thomson Reuters Science Citation Index-Expanded®. Comparisons show that the citations average is the most effective scaling parameter, when the average is based only on the publications actually cited.

# 1. Introduction

The last decade has seen a progressive increase in the use of bibliometric techniques in national research evaluation exercises, which had traditionally been conducted using only peer-review methodology.

For example, in Italy, the former peer-review Triennial Evaluation Exercise (VTR, 2006) has been substituted by a new Assessment of Quality in Research (VQR, 2011), in which panels of experts can choose citation analysis or peer review, or both, for evaluating outputs submitted by universities and research institutions. In the United Kingdom, the previous series of peer-review Research Assessment Exercises (latest, RAE 2008) will be substituted in 2014 by the Research Excellence Framework (REF, 2011), a peer review informed by citation counts and quantitative indicators. In Australia, the most recent Excellence in Research for Australia initiative (ERA, launched 2010) was conducted entirely through a pure bibliometric approach, for the hard sciences: single research outputs were evaluated by a citation index referring to world and Australian benchmarks.

In spite of the advantages of the bibliometric method over peer review in large scale evaluations in the hard sciences, as shown by Abramo and D'Angelo (2011), a number of problems concerning bibliometric applications are still to be resolved. One problem arises from the different timelines of publications and citation behavior in the various research fields, due to the different production functions and coverage of the bibliometric databases for the different fields. The distortions cannot be avoided by simply taking the step of classifying the researchers to be evaluated, according to expertise, and then comparing performance among researchers of the same field. It is well known that individuals often apply their knowledge in transdisciplinary research: we can think of statisticians, who publish in journals in fields of medicine, agricultural science, astronomy, social science, and so on. A classic example in bibliometrics is the case of the physicist Jorge E. Hirsch, who other than being author of numerous articles in physics, is also the inventor of the renowned index known by his name (Hirsch, 2005). This transdisciplinary phenomenon introduces distortions in performance rankings. To deal with this, all bibliometricians agree that it is necessary to carry out so-called field standardization of citations, which are the indicator "par excellence" of the quality of a scientific product.

Standardization involves classifying each article according to its subject category, and then subsequent scaling of the citations in order to render the distributions of citations in each subject category comparable. The scaling is carried out by multiplying the citations of each publication by a factor that characterizes the distribution (for example the inverse of the mean or median) of the citations of articles from the same year and subject category. However, both in literature and in practice, there is still disagreement over the choice of the most effective scaling factor.

At the level of practitioners, the world renowned "crown indicator" (Moed et al., 1995), of the CWTS of Leiden, scales citations of a given publication set with respect to the mean of the distribution. The Karolinska Institute's "field normalized citation score", also uses the mean as scaling factor, appropriately applied to the citations for each single publication (Rehn et al., 2007). However the current authors, observing the strong skewness of the citation distributions, scale the citations to the median of the distribution, for their "Scientific Strength" performance indicator (Abramo et al., 2011).

At the level of the literature, studies support approaches and theses that are often contrasting, but based on empirical tests that are limited to specific fields. As early as 1986, Vinkler suggested the use of the ratio between number of citations and number of papers published in the journals of a whole field as a reference standard for the Relative Subfield Citedness (RW) index (Vinkler, 1986). Radicchi et al. (2008) addressed the problem within 20 fields of several disciplines. Using data from Thomson Scientific's Web of Science of 1999 and 2004, they show that all distributions of different years or fields are rescaled on a universal curve when the average number of citations per article is considered. In the wake of this work, Radicchi and Castellano (2011) provided a deeper study of the fields exclusive to Physics, considering all papers in journals published by the American Physical Society from 1985 to 2009. They confirmed that "when a rescaling procedure by the average is used, it is possible to compare impartially articles across years and fields" and added that "the median is less sensitive to possible extreme events such as the presence of highly cited papers, but dividing the raw number of cites by the median value leads to less fair comparisons and only for sufficiently old publications". In contrast, Lundberg (2007) suggests that due to the strong skewness of distributions of citations, it is preferable to use the median or the geometric mean to scale citations, but he demonstrated that the "item oriented field normalized logarithm-based citation z-score average" (or citation z-score) is better. In keeping with Lundberg, Bormann and Daniel (2009), using a dataset of 1,899 manuscripts submitted to the German chemistry journal Angewandte Chemie International Edition, tested the advantage of using the z-score at micro level, as substitute for normalization to the average. In support of use of the mean as scaling factor, Glanzel (2008) objects to those who think that the application of classical tools of moment-based statistics is not appropriate in research evaluation. In fact, he states that "according the central limit theorem, if the number of observations is large, the distribution of the means of random samples is approximately normal". Still, in the actual practice of evaluation exercises, particularly those not on a particularly large scale, as occurs for some countries, it is not unusual to have subject categories with a low number of publications.

All these previous studies, intended to support decisions on the most effective scaling factor to adopt for evaluation exercises, suffer from two principle limits. The first is that the empirical analyses refer to specific disciplines and the extension of the results to other disciplines is not so readily assumed. The second is that the conditions surrounding the tests have not simulated the typical practices of an evaluation exercise, in which: i) the scale is nation-wide; ii) the period of observation for scientific production is generally five or six years; iii) the date of observation for the citations is close to the last year of the period under evaluation. The aim of our work is to overcome the current limitations, furnishing more accurate and reliable indications regarding the most effective scaling factor for evaluations of all hard science subject categories. The specifics of the tests are formulated with reference to a hypothetical national research assessment exercise. The reference nation chosen is Italy; the organizations are all universities and public research institutions; the years of observation for the scientific production are 2003 and 2007; citations are counted as of 31/12/2008, meaning soon after the period of observation; the disciplines considered are the hard sciences, where bibliometric indicators represent robust proxies to assess research performance. The national scale requires that the effectiveness of the scaling factor be tested on all research disciplines practiced in the country. The cross-time effectiveness is verified for the first and last years of the period of observation, 2003 and 2007, meaning the years that present the maximum and minimum number of citations, counted at the end of 2008.

## 2. Methodology

### 2.1 Dataset

The dataset used for the analysis was extracted from the Italian Observatory of Public Research (ORP - www.orp.researchvalue.it), a database developed by the authors and derived under license from the Thomson Reuters Italian National Citation Report. The ORP contains all the scientific publications authored by scientists from Italian research organizations (95 universities, 76 research institutions and 192 hospitals and health care research organizations), beginning from 2001. The field of observation is limited to the 164 WoS hard science subject categories, grouped into eight disciplines: Biology, Biomedical research, Chemistry, Clinical medicine, Earth and space science, Engineering, Mathematics, Physics. Table 1 shows the scientific production (articles, article reviews and conference proceedings) indexed in the ORP for the years 2003 and 2007, grouped by discipline, composing a total of 94,090 publications.

| Discipline | WoS categories | 2003 | | | 2007 | | |
|---|---|---|---|---|---|---|---|
| | | Public.* | Citations* | Citations per publ. | Public.* | Citations* | Citations per publ. |
| Biology | 29 | 6,970 | 100,209 | 14.4 | 8,916 | 34,611 | 3.9 |
| Biomedical research | 14 | 6,607 | 112,144 | 17.0 | 7,782 | 42,580 | 5.5 |
| Chemistry | 8 | 4,542 | 59,049 | 13.0 | 5,098 | 20,955 | 4.1 |
| Clinical Medicine | 39 | 10,188 | 170,258 | 16.7 | 12,844 | 63,292 | 4.9 |
| Earth and Space Sciences | 12 | 2,725 | 24,767 | 9.1 | 3,469 | 9,094 | 2.6 |
| Engineering | 39 | 10,150 | 53,842 | 5.3 | 13,254 | 17,859 | 1.3 |
| Mathematics | 5 | 2,165 | 9,849 | 4.5 | 2,692 | 3,045 | 1.1 |
| Physics | 18 | 10,075 | 102,771 | 10.2 | 11,821 | 37,990 | 3.2 |
| Total | 164 | 42,067 | 522,536 | 12.4 | 52,023 | 189,596 | 3.6 |

*Table 1: Descriptive statistics for WoS publications, year 2003 and 2007, by discipline; citation count at 31/12/2008.*
*\* Total of values in the column is greater than the figure in the bottom line, due to multiple counts for publications belonging to more than one subject category falling in different disciplines.*

### 2.2 Scaling factors

In the proposed analysis, we compare six scaling factors for citations. We identify the value of standardized citations for each publication[2] as "Article Impact Index" (AII) and formulate six variants of this indicator, one for each scaling factor. In detail, the Article Impact Indexes considered are:

1. $AII_{max} = \frac{c}{|max-min|}$ : ratio between the number of citations for a publication and the range of variation of the distribution of citations for the same WoS subject category and year. Since the minimum value of citations in each distribution is typically nil, this indicator reduces to the ratio to the maximum value.
2. $AII_m = \frac{c}{mean}$ : ratio between the number of citations for a publication and the average value of the distribution of citations for the same WoS subject category and year.
3. $AII_{m0} = \frac{c}{mean\ (no\ zero\ citations)}$: variant of the preceding indicator, with the difference that publications with nil citation value are not included in calculation of the mean.
4. $AII_{box} = \frac{\widetilde{c+1}}{\widetilde{mean}}$: ratio between the number of citations of a publication plus one, transformed by Box-Cox[3], and the average value of such transformed citations for the same year and subject category.
5. $AII_{med} = \frac{c}{median}$: ratio between the number of citations for a publication and the median value of the distribution of citations for the same year and subject category. This indicator cannot be calculated in cases where more than 50% of the publications of a WoS category have zero citations.
6. $AII_{med0} = \frac{c}{median\ (no\ zero\ citations)}$: variant of the preceding indicator, with the difference that publications with nil citation value are not included in calculation of the median.

---
[2] In the current analysis, the authors do not distinguish publications by type.

## 3. Results and analysis

### 3.1 Distribution of citations

The literature notes that citations distribution is very skewed to the right: this means that most papers are relatively little cited and there are only a few papers with many citations. In Figure 1, as an example, we show the distributions of citations (counted on 31st December, 2008) for Italian publications in 2003 and 2007 in the WoS category Oncology, Biomedical research discipline.

[Figure 1]

The graphed distributions, with their long tails to the right, are very far from Gaussian form. Moreover although form of the distribution curves are the same, they show different statistical parameters, such as different values of mean and median. Various works have dealt with the study of such distributions. According to some authors, citation distribution follows a power law, characterized by a rapid decrease in frequency of citations beyond a certain threshold (Peterson et al., 2010; Albarrán et al., 2009; Gupta et al., 2005), while others state it as well represented by a double exponential-Poisson distribution for all categories (Vieira and Gomes, 2010). In spite of the high skewness, attempts have been made to adjust citation distribution to normal; Lundberg (2007), in particular, adjusted the distribution to Gaussian with nil mean and variance of one. First he normalized the distribution by a natural log function, and then standardized citations to the "z-score". The results were apparently reassuring, however they were obtained over a few well delineated categories (Cell Biology, Biochemistry and Molecular Biology, Clinical Neurology and Crystallography) and considering a very long citation window, which are conditions not found in a true evaluation exercise. For reasons of space, we provide only the descriptive statistics for the 14 WoS categories of the Biomedical research discipline, referring to publications by researchers in Italy, years 2003 and 2007 (Table 2).

---

[3] Box-Cox transformation reduces asymmetry and reshapes the distribution as close as possible to Gaussian, through applying a λ parameter (usually between -3 and +3) estimated by the likelihood method. We added one to citation values to take into account publications with no citations.

| Subject category | Year | Public. | Mean | Median | St.dev. | Skewnees | Kurtosis |
|---|---|---|---|---|---|---|---|
| Allergy | 2003 | 115 | 18.24 | 11 | 21.58 | 2.45 | 6.91 |
|  | 2007 | 163 | 5.37 | 3 | 7.43 | 3.25 | 15.37 |
| Anatomy and morphology | 2003 | 93 | 9.18 | 5 | 22.73 | 7.74 | 66.74 |
|  | 2007 | 89 | 1.91 | 1 | 2.66 | 2.20 | 5.32 |
| Oncology | 2003 | 1,730 | 17.48 | 9 | 27.99 | 4.51 | 29.52 |
|  | 2007 | 1,927 | 6.38 | 3 | 15.14 | 10.81 | 172.56 |
| Chemistry, medicinal | 2003 | 344 | 15.75 | 10 | 26.98 | 7.39 | 72.70 |
|  | 2007 | 579 | 4.54 | 3 | 5.55 | 3.36 | 19.96 |
| Hematology | 2003 | 789 | 26.58 | 13 | 42.33 | 4.54 | 30.81 |
|  | 2007 | 836 | 8.99 | 4 | 13.58 | 3.60 | 18.42 |
| Immunology | 2003 | 941 | 20.49 | 11 | 36.95 | 9.90 | 163.77 |
|  | 2007 | 1,071 | 6.84 | 3 | 13.99 | 7.10 | 66.8 |
| Infectious diseases | 2003 | 372 | 14.14 | 9 | 19.48 | 5.51 | 54.36 |
|  | 2007 | 371 | 4.93 | 3 | 8.12 | 5.49 | 41.65 |
| Medical laboratory technology | 2003 | 145 | 11.90 | 7 | 23.29 | 5.43 | 35.29 |
|  | 2007 | 195 | 2.92 | 2 | 3.42 | 2.29 | 7.62 |
| Medicine, research and experimental | 2003 | 568 | 17.81 | 6.5 | 37.56 | 5.98 | 48.13 |
|  | 2007 | 513 | 6.60 | 3 | 15.88 | 6.28 | 49.47 |
| Pathology | 2003 | 365 | 12.20 | 7 | 14.92 | 2.82 | 11.48 |
|  | 2007 | 435 | 3.80 | 2 | 5.89 | 5.58 | 51.57 |
| Pharmacology and pharmacy | 2003 | 1,261 | 14.44 | 9 | 21.22 | 6.39 | 66.43 |
|  | 2007 | 1,584 | 4.51 | 3 | 5.88 | 4.31 | 36.97 |
| Radiology, nuclear medicine & med. imaging | 2003 | 813 | 7.81 | 1 | 19.27 | 9.21 | 125.04 |
|  | 2007 | 981 | 2.30 | 1 | 4.00 | 4.06 | 28.18 |
| Toxicology | 2003 | 287 | 14.47 | 9 | 18.97 | 3.78 | 21.35 |
|  | 2007 | 376 | 4.08 | 2 | 5.45 | 3.14 | 12.77 |
| Virology | 2003 | 227 | 17.59 | 11 | 22.16 | 5.54 | 49.73 |
|  | 2007 | 244 | 5.44 | 3 | 7.86 | 4.07 | 20.98 |

*Table 2: Descriptive statistics of citation distributions for publications in the 14 subject categories of Biomedical Research, years 2003 and 2007*

We observe that average and mean values are very different across categories or, within the same category, across years of publication. The 2003 publications in Oncology, observed as of 31/12/2008, receive a mean 17.48 citations, with a median equal to 9. These values diminish notably for the 2007 publications: average value drops to 6.38 citations, median to 3. Distribution of the 2007 publications is much more asymmetric (skewness: +10.81). Overall, the highest average and median citation values are seen for the publications from 2003 in Hematology, respectively at 26.58 and 13; while the lowest average value is seen for 2003 publications in Radiology, Nuclear medicine and medical engineering category, at 2.30. A detail of this particular WoS category is that both year cohorts examined have a median value of 1.

Descriptive statistics were repeated for each of the WoS hard science categories, always showing the same type of distribution with long tail to the right, but with very different statistical parameters across categories. To compare the categories, we calculate the probability of observing a number of citations greater than or equal to number of citations "c". In this manner it is possible, in a single graph, to plot very different frequency distribution. Figure 2 presents the log scale distributions of these probability distributions, for all 164 WoS categories in the eight hard sciences.

[Figure 2]

We observe clearly that, when we consider different years and categories, the citation distributions are incompatible. It is not at all possible to superimpose the probability curves, and with these conditions it is impossible to carry out a comparison between categories, without an adequate rescale operation.

## 3.2 Analysis of standardized distributions

*Standardization to maximum value*

Our first standardization test is with indicator $AII_{max}$, meaning the maximum value of each distribution of citations. In general, this method of standardization would be affected by the presence of anomalous citation values. In Figure 3, as an example, we present box plots of the distributions of the Biomedical research WoS categories. One can clearly observe, for example, that within the Immunology (NI) category, for the year 2003, the maximum value (citations = 740) is an outlier and cannot be used as benchmark for all other citation values, since this would underestimate each standardized values.

[Figure 3]

The significant presence of anomalous values within such distributions leads the authors to set aside this standardization methodology and not carry out any further empirical verification.

*Standardization to mean value*

One of the most frequently adopted solutions is to standardize the number of citations for a publication to the average value of citations for the publications of the same WoS category and year of publication. In calculating the mean, the choice can be made as to whether or not to include the publications that do not receive any citations. We consider both options, thus using indicators $AII_m$ and $AII_{m0}$. In Figure 4 we present the probability curves for $AII_m$ per year of publication, for the 164 categories analyzed. We observe that such standardization produces results that are not optimal for more recent publications, near to the date of the citations count (2008). In fact, for 2007 publications, the indications are of limited capacity for scaling and a divergence of the probability curves for some categories. However for the publications of longer date, such as 2003, the results are more comforting, though still not optimal, since some categories dissociate in evident fashion.

As for inter-category comparison, as logically expected, the superimposition of scaled curves for inter-temporal examination is also not optimal.

[Figure 4]

The same operation was carried out for values of $AII_{m0}$, meaning without taking account of publications that receive nil citations. In Figure 5 we observe that this standardization produces better results than the preceding one, both for publications from 2003 and from 2007.

[Figure 5]

Figure 6 groups the distributions of both the 2003 and 2007 cohorts in a single graph: the inter-temporal comparison also seems almost optimal except for the extreme values, in which the curves seem to diverge.

[Figure 6]

*Standardization to mean after Box-Cox transformation*

Literature notes the mean as an optimal estimator of distribution central tendency, provided that the distribution is symmetrical, but this is a characteristic which does not occur in distributions of citations. For this reason, there have been attempts to transform the citation distributions to Gaussian in order to reduce asymmetry to a minimum, and then normalize the transformed values to the mean. Applying the Box and Cox function (1964), we transform each citation value in function of parameter $\lambda$, estimated through the method of maximum likelihood, as follows:

$\frac{(c+1)^\lambda - 1}{\lambda}$ per $\lambda \neq 0$; $log\ (c+1)$ per $\lambda = 0$

In this case, values of $\lambda$ vary from a minimum of -3.000 for WoS category "Imaging science and photographic technology", to a maximum of +0.717 in "Integrative and Complementary Medicine".

Figure 7 presents the graphic for the standardization of Box-Cox transformed citations of 2003 publications. The reduction of asymmetry leads to an extreme compression of citation values and a notable reduction in the range of variation; nevertheless, the result is that the standardized curves do not superimpose.

[Figure 7]

*Standardization to median values*

Figure 8 presents the distributions as standardized with respect to the median value of the original distributions, subdivided by year and subject category (*AII$_{med}$*). For 2003, only 139 categories can be represented, having a median citation value other than zero. For 2007, the number of WoS categories represented drops to 121. In any case, we observe that the median offers better capacity for scaling for the most recent publications (from 2007), although not optimal, and with presence of one (PY-General and internal medicine) that diverges from the others in very evident manner.

[Figure 8]

Graphing the two cohorts of publications (Figure 9), we see a limited capacity for this method of standardization to render publications from different years comparable.

[Figure 9]

When we exclude the zeros from the calculation of median values of distributions, the results improve. Still, for *AII$_{med0}$*, the scaling is better for publications that are more remote (2003) than for recent ones (2007). In Figure 10 we see that for the 2003 publications, the standardized values start to diverge beginning from a certain number of citations, while the 2007 values already diverge at the base of the distribution.

[Figure 10]

The inter-temporal comparison converges only up to the point of a number of standardized citations equal to about one ($\log AII_{med0} = 0$).

In the analyses conducted above we see various biases in the different types of standardization, especially in the tails of the distributions, for the so-called extreme values. For this reason we conducted a further analysis focused on the top publications in terms of citations. We would expect that an effective standardization would guarantee a constant share among the global top publications for all categories and years. The results are presented in the next section.

### 3.3 Analysis of top publications

We first compute a global ranking of all hard sciences publications, on the basis of citations received. Next we extract the publications that enter in the global top 10%. If the distribution of the citations were not to differ across subject categories, the expected percentage of top papers would be around 10% in each category. Admissible variation of this percentage is measured by standard deviation, in keeping with Radicchi and Castellano

(2011).

Figure 11 shows the percentage of papers belonging to the top 10% of the global ranking per non-standardized citations, for the 164 categories analyzed.

We observe that these percentages fluctuate greatly with change in category. In fact, for Medicine, General and Internal more than 30% of publications are part of the global top 10% of rankings, while for 12 out of the 164 WoS categories there are no publications in top. In general, just 33 out of 164 WoS categories show a range of percentages of top that fall in the $10\% \pm 1\, s.d.$ interval.

[Figure 11]

This operation, repeated for citations standardized to mean and to median under various alternative described, gives the results presented in Figure 12. The percentages shown in the two upper quadrants show the two standardizations to mean: (a) is with the calculation including nil-cited publications and (b) is without inclusion of the nil-cited publications.

In the first case, the categories are well represented in the global top 10%, where 102 WoS categories out of 164 place between the upper and lower bounds, as defined. In the second case, this count further increases to 114 (about 70%).

In contrast, in the lower quadrants, both the standardizations to median including the nils (c) and excluding the nils (d) show a limited capacity for this method of standardization to filter the specificity of each subject category. The greater part of the categories place outside the interval of admissibility and the fluctuations between categories are truly notable. For case (c), or standardization relative to the median with inclusion of nils, only 28 (out of 139 categories where the median of citations is greater than zero) present percentages of global top that fall within the interval. For case (d), 68 out of 164 categories show non-distorted percentages.

[Figure 12]

**Discussion and conclusions**

The problem of scaling becomes especially critical when comparative evaluations must consider publications from different years and subject categories, as is typical in national research evaluation exercise. All published studies on the most effective scaling factor have suffered from clear limits, particularly because the experimentation has not reflected the conditions of the true exercises. This work attempts to resolve this critical problem and thus provide research assessment practitioners with accurate and reliable indications.
The study tests six scaling factors on the dataset for a full hypothetical national research assessment concerning all hard science fields. The examination shows that the best results in comparability of standardized impact of publications over different years and subject categories are obtained by scaling the citations to the average value of their relative distributions (with nil values removed). This scaling factor is not as effective for the extreme values of citation distributions (top publications) as it is for other regions however, in accordance with Radicchi and Castellano. (2008) and Radicchi et al. (2011) even here it is certainly more effective than all the other standardization methods tested.

Few findings from our analysis are not aligned withthose by Radicchi et al. (2011), whose field of osbervation referred to all world articles indexed in WoS(cohorts of 1999 and 2004) in 20 subject categories. They have demonstrated that the distribution of standardized citations to mean fits with a lognormal curve for all categories considered, for a standardized value ≥0.1. On the contrary, our findings show that the citation distributions of Italian publications do not fit with that curve. For example, the distribution of standardized citations to mean regarding all Italian publications of 2004 classified in the subject category Hematology (MA), presents a value of Anderson-Darling goodness of fit test for the lognormal distribution equal to 3.291 (p-value < 0.005), for Article Impact Index ≥0.1. Furthermore, they state that the inclusion of uncited articles produce just a small shift in mean values. Our analysis instead leads to different conclusions. For example, for the 2004 Italian publications in Neuroimaging (RX), the mean of citations of all publications is 4.14, while the mean calculated without including uncited publications increases to 13.83 (+234%). Differences in the conclusions may be partly explained by the fact that Radicchi et al. consider all world articles, while we consider Italian publications only; furthermore, we include also conference proceedings in the analysis, which notoriously are less cited than articles. Therefore, the decision to include or not uncited publications in the distribution when calculating the scaling factor, is very critical when citation distributions refer to a national scale or to different types of publications.

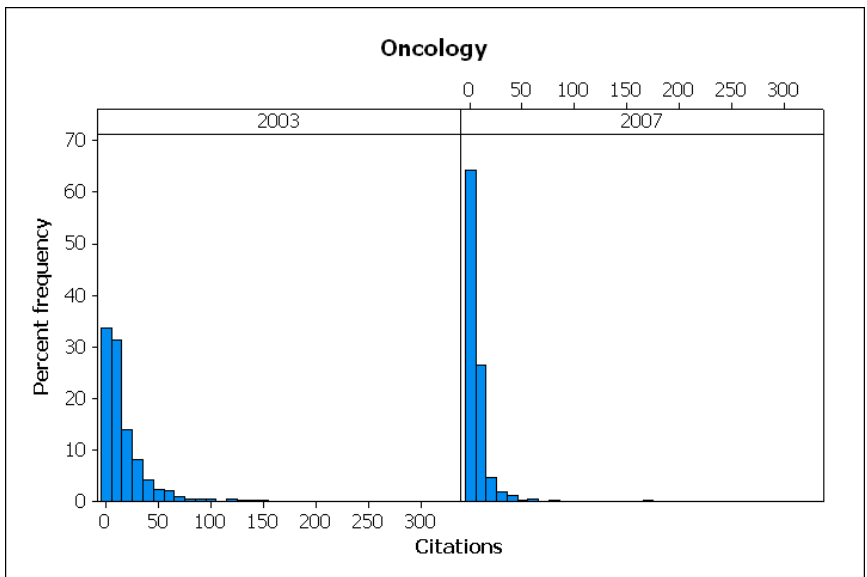

*Figure 1: Citation distributions of 2003 and 2007 publications in Oncology*

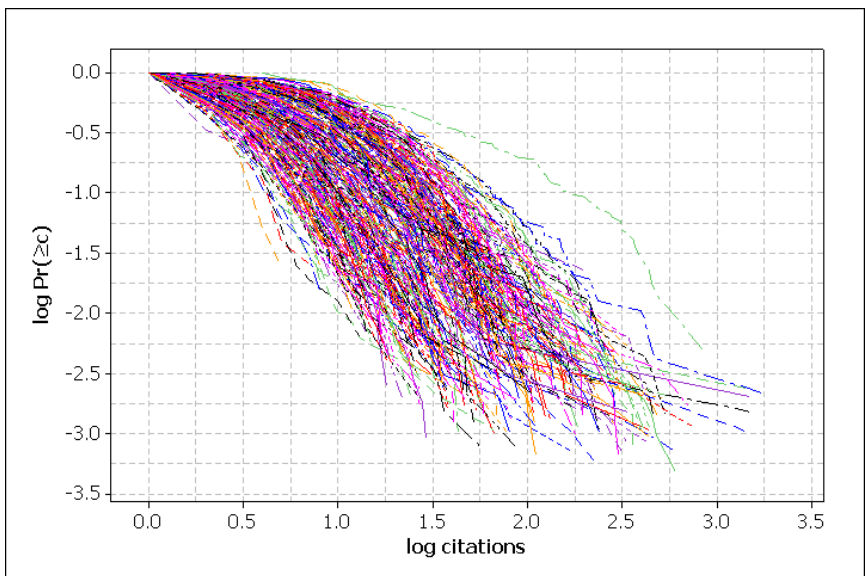

*Figure 2:* **Probability Pr(≥c) to observe a paper with more than (or equal to) c citations by category for the 2003 and 2007 publications.**

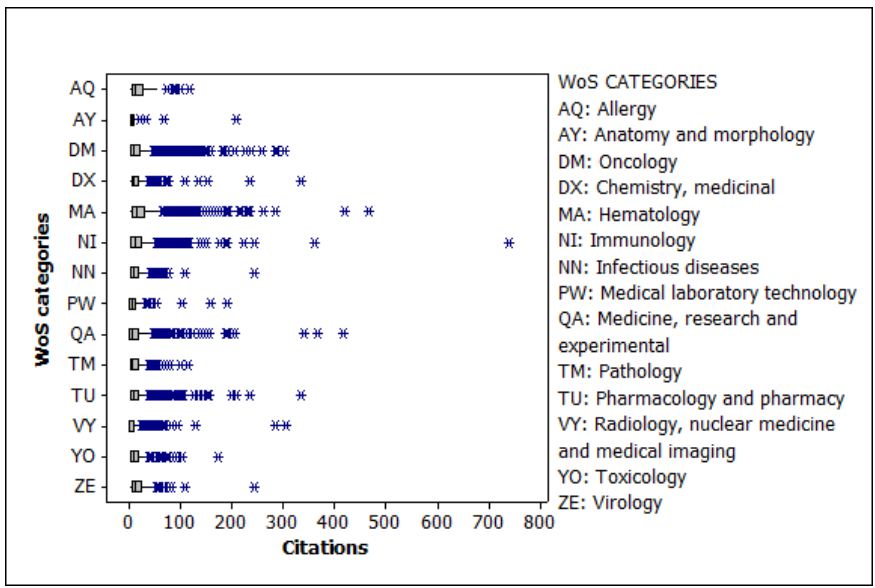

*Figure 3: Box plot of citations for Biomedical research WoS categories (publication year: 2003).*

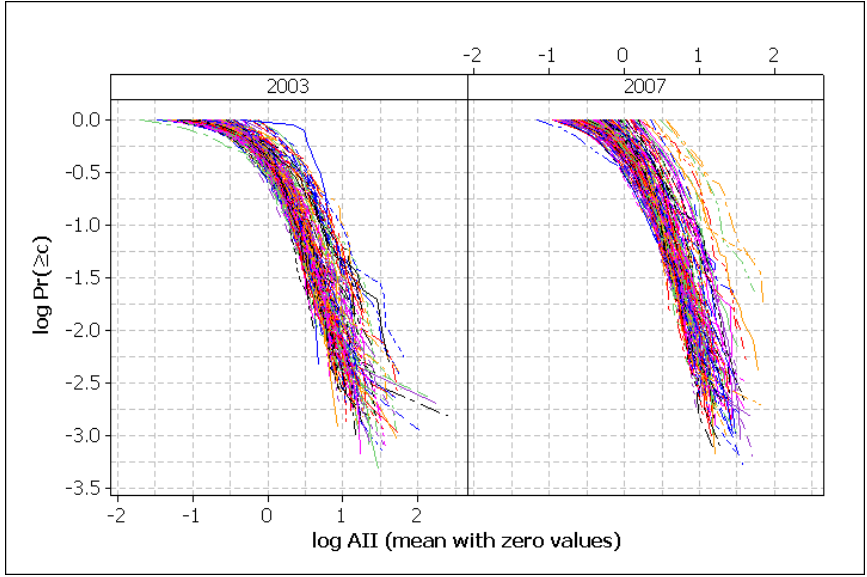

*Figure 4: Probability P(≥c) of observing a paper with greater than or equal to "c" citations, standardized to mean by category (calculated including nil-cited publications), for publications from 2003 and 2007.*

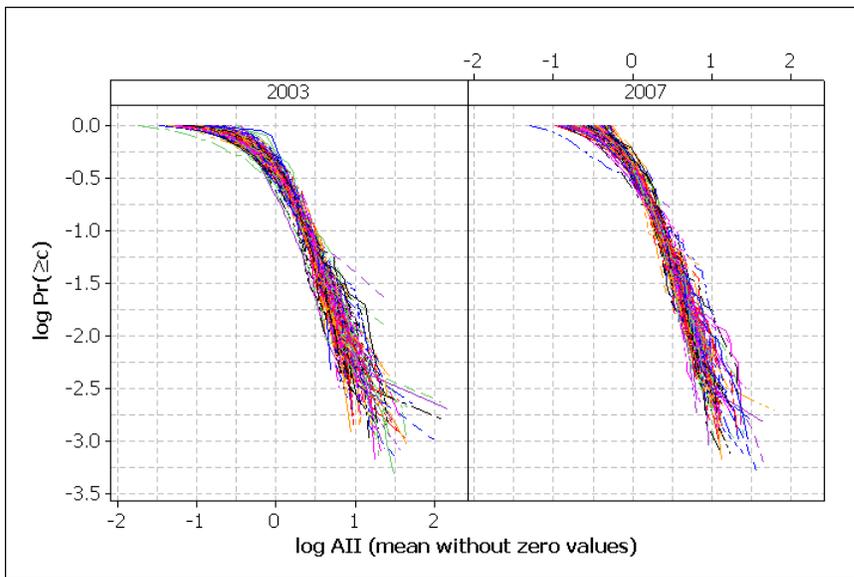

*Figure 5: Probability P(≥c) of observing a paper with greater than or equal to "c" citations, standardized to mean by category and by publication year (calculated without including nil-cited publications).*

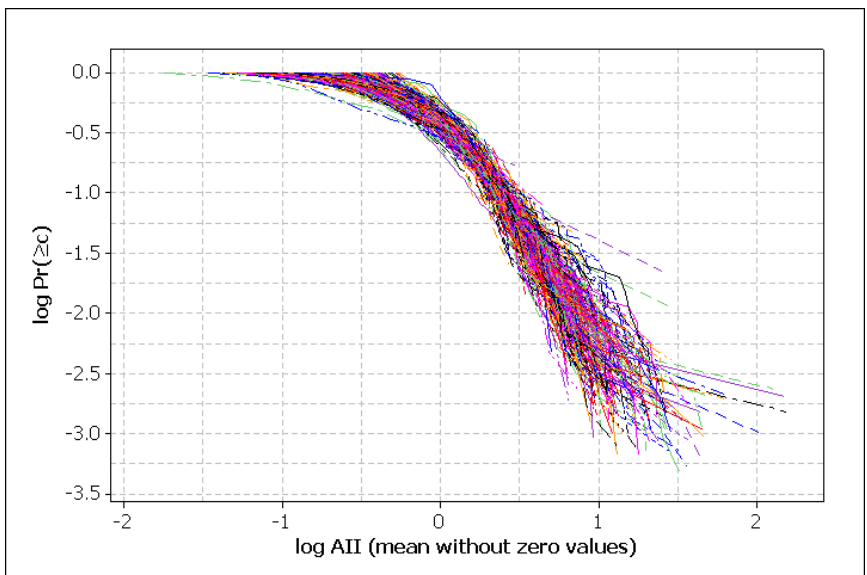

*Figure 6: Probability P(≥c) of observing a paper with greater than or equal to "c" citations, standardized to mean by category (calculated without including nil-cited publications), for publications form 2003 and 2007.*

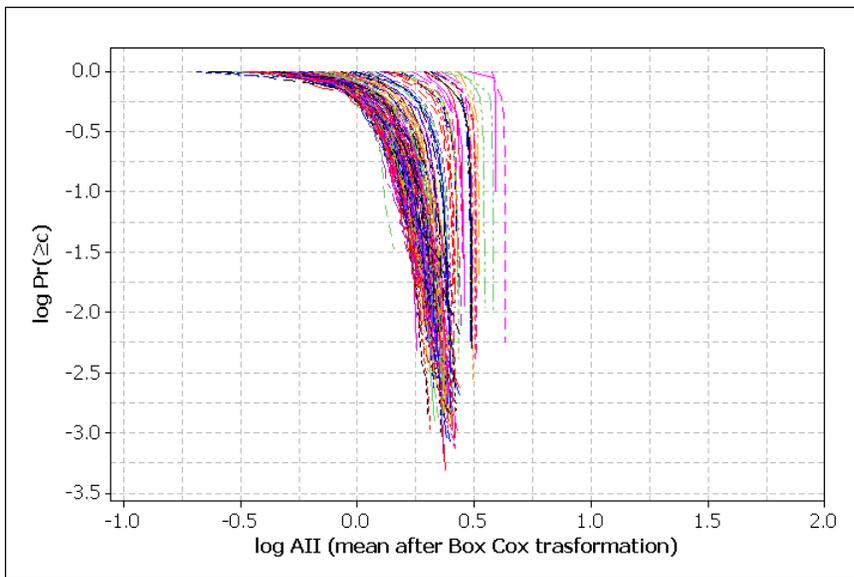

*Figure 7: Probability P(≥c) of observing a paper with greater than or equal to "c" citations, standardized to mean by category after Box-Cox transformation, for publications from 2003.*

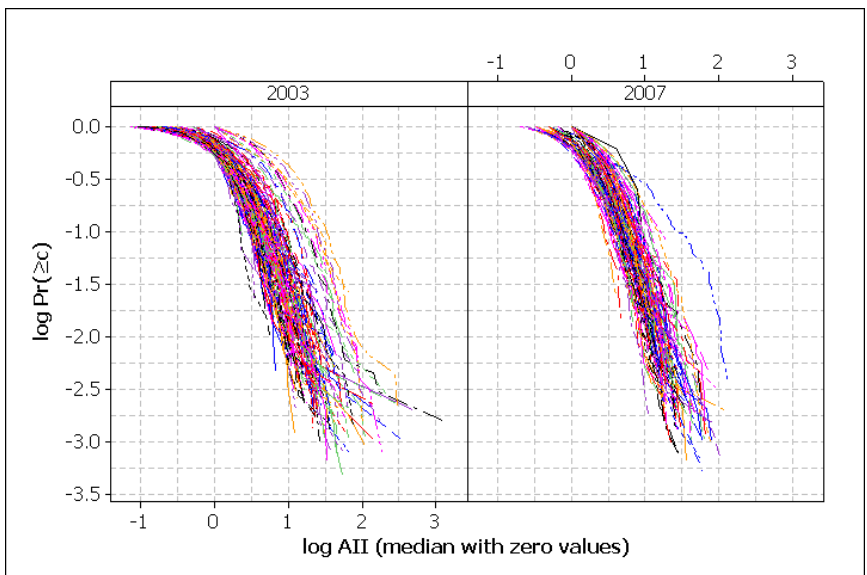

*Figure 8: Probability P(≥c) of observing a paper with greater than or equal to "c" citations, standardized to median by category (calculated including nil-cited publications), for publications form 2003 and 2007.*

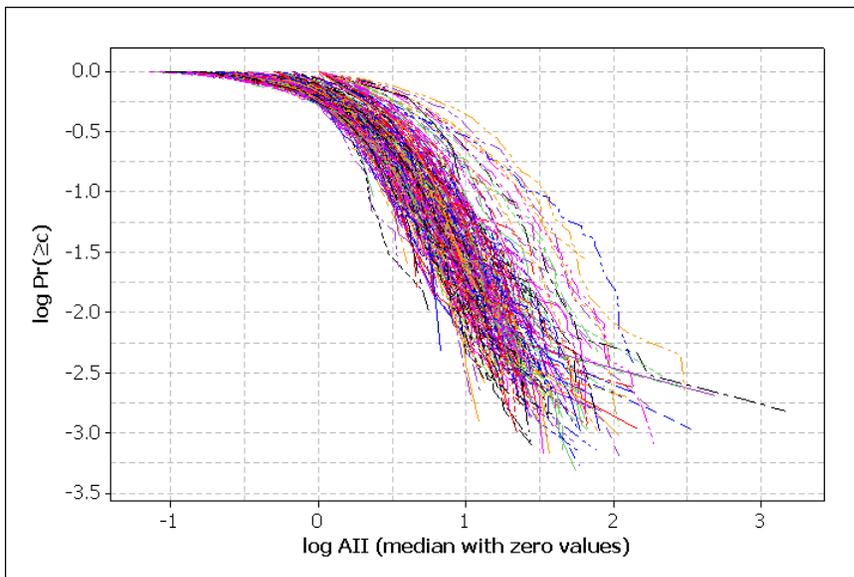

*Figure 9: Probability P(≥c) of observing a paper with greater than or equal to "c" citations, standardized to median by category (calculated including nil-cited publications), for publications form 2003 and 2007.*

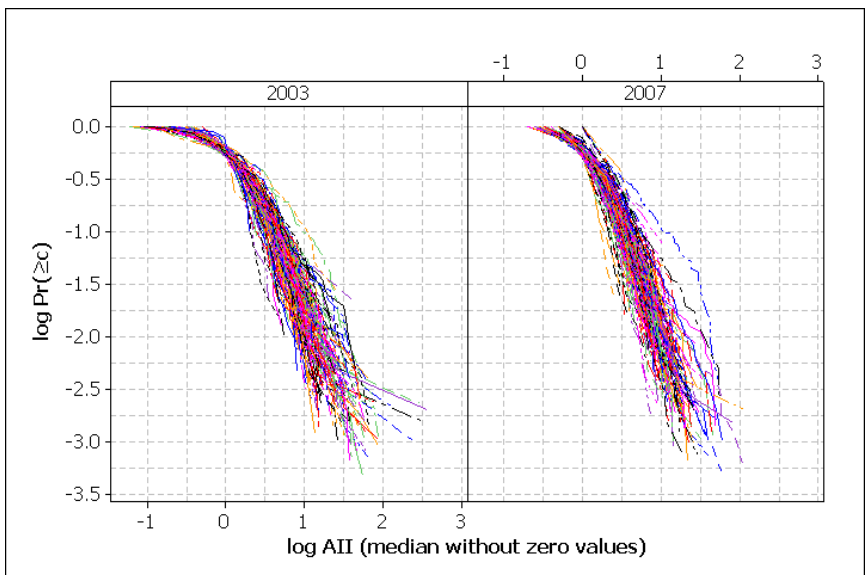

*Figure 10: Probability P(≥c) of observing a paper with greater than or equal to "c" citations, standardized to median by category and publication year (calculated without including nil-cited publications).*

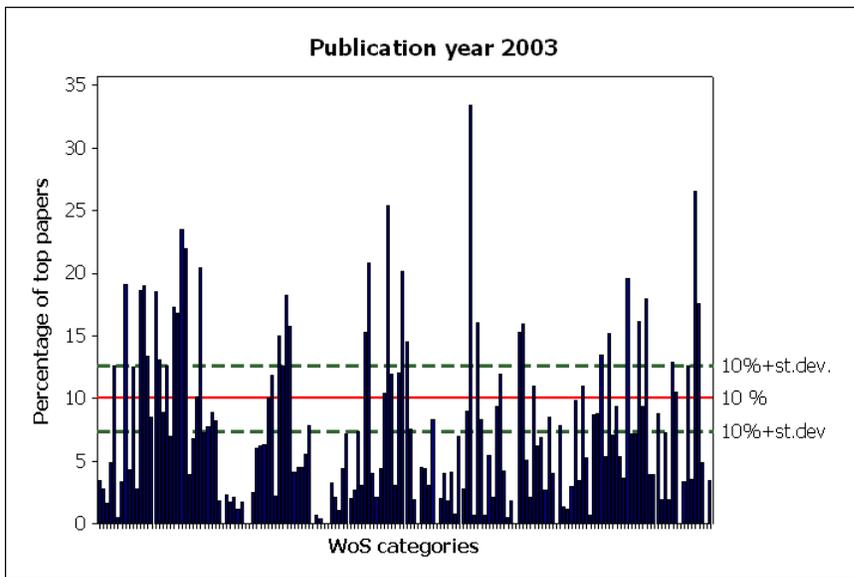

*Figure 11: Percentage of papers belonging to the top 10% of the global ranking according to citations by WoS category.*

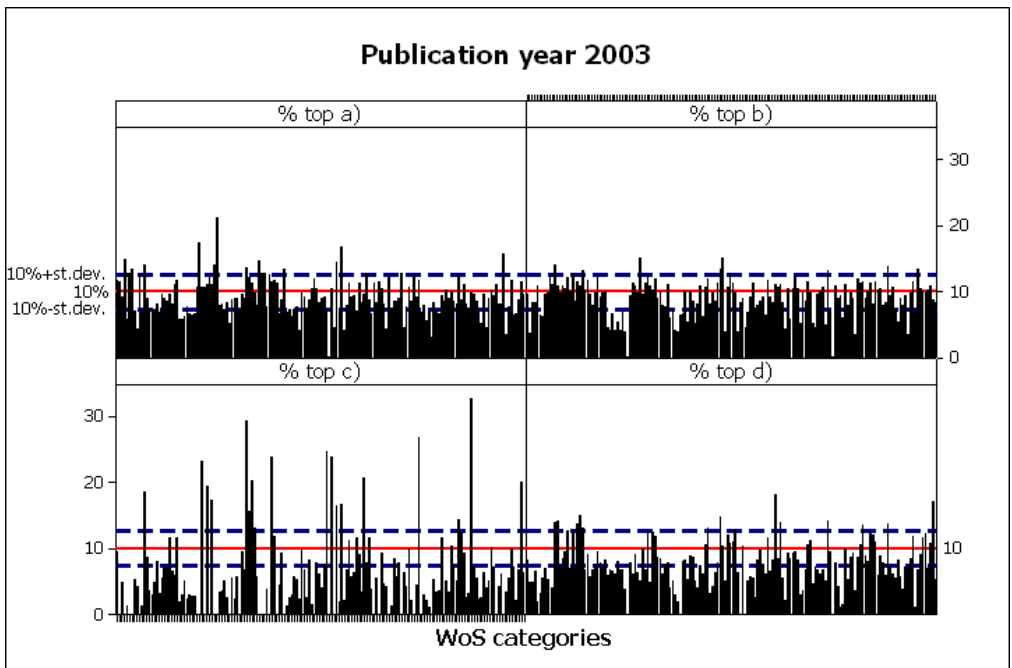

*Figure 12: Percentage of papers belonging to the top 10% of the global ranking by WoS category and type of standardization - a) Standardization to mean value witht zero values; b) Standardization to mean value without zero values; c) Standardization to median value with zero values; d) Standardization to median value without zero values.*